\magnification \magstep1
\centerline {\bf Renormalization of Wilson Operators in Minkowski space}
\vskip 2cm
\centerline  { A. Andra\v si \footnote{$^{\dagger}$}{andrasi@thphys.irb.hr} }
\centerline {\it `Rudjer Bo\v skovi\' c' Institute, Zagreb, Croatia }
\vskip 1cm
\centerline { J.C. Taylor \footnote{$^*$}{jct@damtp.cambridge.ac.uk}}
\centerline {\it Department of Applied Mathematics and Theoretical Physics,}
\centerline {\it University of Cambridge, Cambridge, UK }
\vskip 4cm
 We make some comments on the renormalization of Wilson operators
 (not just vacuum -expectation values of Wilson operators),
 and the features which arise
in Minkowski space.  If the Wilson loop contains a straight light-like segment,
 charge renormalization does not work in a simple graph-by-graph way;
but does work when certain graphs are added together. We also verify that, in a
 simple example of a smooth loop in Minkowski space, the existence of
pairs of points which are light-like separated does not cause any extra
divergences.

\vfill \eject
    
\beginsection  1. Introduction

There has been some discussion in the literature [1-8] of the
renormalization of Wilson operators, but some of this has been
done in Euclidean space [1,3,6], and most has been restricted to
the vacuum-expectation value of the Wilson operator rather than
the operator itself. In [2], we have previously noted some
complications which appear in the renormalization of the Wilson
operator in Minkowski space, when the Wilson loop contains a
light-like segment. In particular, we noted that individual graphs
do not always have divergent parts which are local. Here we show
that renormalization nevertheless works provided that suitable
sets of graphs are taken together.

Let the operator be
$$ W = tr P \exp g\int_C A.dx \eqno(1)$$
where $C$ is a closed curve, $P$ denotes operator and matrix ordering
along $C$, and the nonabelian gauge field $A_{\mu}$ is a matrix in
some representation $R$ of the gauge group $G$. Sometimes one is interested
just in the vacuum-expectation value
$$ \langle W \rangle_. \eqno(2)$$

$W$ is an example of a composite operator, and one hopes that it is
multiplicatively renormalizable, in the sense that
$$ W_R (A_R;g_R) = Z(\epsilon)W_B(A_B;g_B,\epsilon), \eqno(3)$$
where suffices $R$ and $B$ denote renormalized and bare quantities.
Dimensional regularization is used, with $d=4-\epsilon$, and the 
presence of $\epsilon$ in some terms in (3) signals that they are
divergent. The relationship between $g_B$ and $g_R$ and between $A_R$
and $A_B$ should be the same as in ordinary perturbation theory.

If we take the vacuum-expectation value of (3), we get
$$\langle W_R(g_R)\rangle =Z(\epsilon) \langle W_B(g_B, \epsilon) \rangle.
\eqno(4)$$
Without any further restriction, (4) is  not very rerstrictive, since it
serves merely to define $Z$.  However, it is known [6,7,8] that
the contributions to $Z$ are associated with geometrical features of $C$,
especially corners and cusps. But, once $Z$ is determined from (4),
condition (3) is certainly nontrivial.

The heuristic reason one expects (3) to hold is that the divergences
are expected to be
short distance features, and therefore local ones on the curve $C$.
For example, the graphs in Fig 1(a) and (b) look like ordinary
 charge-renormaization graphs, and one
 expects the divergences to come from the regions
in which the points $x, y, z$ are close together;
and so these divergences can be cancelled by the charge- renormalization
counter- term graph (c).
In fact the divergences from the subgraphs in (a) and (b) are the same as those
in the ordinary Feynman graphs which they superficially resemble
(with the line denoting the loop being replaced by, say, a quark line).
But if the loop $C$ contains a straight lightlike segment this simple
expectation is not fulfilled, as we discuss in the next section.
\beginsection 2. Straight lightlike segments

Take a Wilson loop which contains a straight lightlike segment $AB$, as in 
Fig.2, where the vector $V$ represents that segment. $A'A$ and $BB'$
are neighbouring portions of the loop.

We use light-cone coordinates defined as follows. Let $V'$ be another
lightlike vector satisfying $V.V' =1$ and with the momenta $k$ and $p$
in Fig.2. Let
$$V.k =u,~~V'.k =v,~~ V.p =w,~~V.K=V'.K=0,\eqno(5)$$
so that
$$k_{\mu}=vV_{\mu}+uV'_{\mu}+K_{\mu}.\eqno(6)$$

We will be concerned with divergences coming from the region where
$u$ is finite and $v$ and $K$ are large with $K^2$ scaling like
$v$. We will assume that this is the case, and verify for in an
Appendix that it is in
a special case. In this spirit, to identify divegences, we neglect all
components of $p$ except $w$ (since this appears along with $u$).
Then Fig.2(b) gives (we use the Feynman gauge)
$$ {1\over 2}ig^3C_G t_aV_{\mu}
\int_0^1ds \int_s^1 dt\int {{d^2K du dv}\over{(2\pi)^4}}e^{-i[(w-u)s+ut]}
 {[(w-u)-u] \over (uv-K^2+i\eta )(v(u-w)-K^2+i\eta)},\eqno(7)$$
where $C_G$ is the gluon colour Casimir and $t_a$ is the colour matrix
in the representation used in the definition of the Wilson loop.
We will treat separately the two terms $(w-u)$ and $-u$ in the
square bracket in the numerator, taking the $-u$ term  first.
Then, carrying out the $t$-integration, we get a factor
$$ e^{-iws} - e^{-iu+i(u-w)s}. \eqno(8)$$
The contributions from these two terms have the structures shown in Fig.3(a)
and (b).
Fig.3(a) contributes half of the usual nonabelian part of the quark
vertex part renormalization
constant (with dimensional regularization with $d=4-\epsilon$),
$$ Z_{1f}=  1 -{2\over{\epsilon}}C_G{{g^2}\over{16\pi^2}} \eqno(9) $$
(in the notation of Itzykson and Zuber eq.(12-122), also $ Z_3 $ is
defined in eq.(12-144)),
the other half coming from the $(w-u)$ term in the numerator of (7), as we shall
discuss below. Together with the gluon vacuum polarization graph
renormalization constant  $ Z_3 $
this is absorbed by coupling constant renormalization.
But the second contribution from (8) inserted into (7),
 denoted by Fig.3(b), is also divergent. We will show that its divergence
is cancelled by the contributions from Fig.2(c) and (d).
Take first Fig.2(c). The numerator contains the factor
$$2V.dx k_{\mu} -k.dx V_{\mu} - V.kdx_{\mu}, \eqno(10)$$
where factors containing $p$ have been neglected and
$dx_{\mu} = {dx_{\mu} \over dr}dr$ is an element of the portion $BB'$
of the loop (parametrized by $r$ with $r=0$ at $B$). The last term in
(10) may also be neglected since $V.k=u$ is not large.

In the contribution from the second term in (10), the integral
along $BB'$ has the form
$$ \int_0 dr~ k.{dx \over dr} e^{-ik.x(r)}, \eqno(11)$$
and the contribution from the lower limit is the relevant one (any other
contribution being finite because of the oscillating exponential).
This contribution has the same structure as Fig.3(b), and it is
easily verified to cancel that term (i.e. the contribution to (7)
from the $ (-u) $ in the numerator).

Next take Fig.2(d), which we shall treat along with the contribution
to Fig. 2(c) from the first term
in (10). Fig.2(d) gives
$$ {1\over 2}g^3C_G t_a V_{\mu}(2\pi)^{-4}
\int dudvd^2K\int dr V.{dx \over dr} e^{-ik.x(r)}\int_0^1 ds \int_s^1 dt
{e^{-iwt+ius} \over uv-K^2+i\eta}$$
$$= -{1\over 2}ig^3C_G t_a V_{\mu}(2\pi)^{-4}
\int du dv d^2K\int dr V.{dx \over dr} e^{-ik.x}\int_0^1 ds {e^{ius}(e^{-iws}-e^{-iw})
\over w(uv-K^2+i\eta)}. \eqno(12)$$
The contribution to Fig.2(c) from the first term in (10) has an $r$-integral
of the same form as in (12).

In order to proceed we need to discuss the curve $BB'$ in more detail.
The factor $e^{-ik.x}$ will make (12) converge except infintesimally close
to $r=0$. Assume that
$$ x_{\mu}(r) =V_{\mu}+c_{\mu}r+O(r^2). \eqno(13)$$
We will consider two possibilities:- (i) That $c_{\mu}\propto V_{\mu}$
so that the tangent to the curve $C$ varies continuously at $B$.
(ii) That $c_{\mu}$ is not proportional to $V_{\mu}$, so that there
is a discontinuity in the tangent to $C$ at $B$.

In case (ii), we may  use in (12) the approximation $ c.k \approx vc.V $
(from (6)); so that
$$V.{dx \over dr} \approx v^{-1}k.{dx \over dr},
 \eqno(14)$$
with neglect of convergent terms. Then the $r$-integration in (12) is exact, and the
contribution from the lower limit $r=0$ is
$$ -{1\over 2}g^3C_Gt_aV_{\mu}(2\pi)^{-4}
\int du dv d^2K \int_0^1 ds {e^{-i(1-s)u}(e^{-iws}-e^{-iw})\over
wv(uv-K^2+i\eta)}.\eqno(15)$$
In the contribution from the second term in the numerator (15)
we make the change of variables $u\rightarrow (u-w)$. Then (15)
becomes
$$ {1\over 2}g^3C_Gt_aV_{\mu} \int {{du dvd^2K}\over{(2\pi)^4}}
 \int_0^1 ds {e^{-iu-is(w-u)} \over vw}\left [{1\over
v(u-w)-K^2+i\eta} -{1\over uv-K^2+i\eta} \right ]. \eqno(16)$$
This cancels the contribution to Fig. 2(c) from the first term in (10)
(since the two denominators in (16) are the same as the Feynman denominators
from Fig. 2(c)).

Next we take the alternative (i) above. In this case we can
suppose that
$$x=V +\alpha rV + {1\over 2} r^2 V'+O(r^3), \eqno(17)$$
where $\alpha$ is a constant.
Then
$$k.x =(1+\alpha r)u+{1 \over 2}vr^2+O(r^3)~,~~ V.dx = rdr+O(r^2). \eqno(18)$$
The relevant range of the $r$-integration is $r\simeq v^{-{1\over 2}}$, so
we may make the approximation
$$k.{{dx}\over{dr}}\approx \alpha u + rv +O(r^2), \eqno(19)$$
with neglect of terms down by a factor $v^{-1/2}$, such terms being therefore
not divergent.
With the use of (19), the $r$-integration again is exact
and the result reduces to  case (ii) previously
considered.

The contribution to Fig.2(b) from the term $w-u$ in the numerator of (7)
is similarly cancelled by graphs like Fig. 2(c) and (d) but with the
gluon $k$ attached to $AA'$ instead of $BB'$.

\beginsection 3. An example of a Wilson loop in Minkowski space

The unexpected divergences encountered in Section 2 come from points on the
curve $C$ which have a null interval between them. For any closed smooth curve,
at a point where the tangent is light-like,
the light cone  intersects the curve again in
at least two other points. Are there similar unexpected divergences connected with
all pairs of points with a null interval between them? In order to
investigate this question, we take the simple case where $C$ is the
ellipse
$$x=(a\cos u ;0,0,b\sin u)~~~~(0 \leq u <2\pi). \eqno(20)$$

In this case, the second order contribution to $\langle W \rangle$ is
$$2g^{2} C_R (2\pi)^{2-\epsilon/2}e^{i\pi \epsilon}\Gamma(1-{1\over 2}\epsilon)
\int_0^{2\pi}du \int_0^{2\pi}dv(a^2\sin u \sin v -b^2 \cos u \cos v)$$
$$\times [a^2(\cos u -\cos v)^2 -b^2(\sin u -\sin v)^2-i\eta]^{-1+\epsilon/2}$$
$$=2g^2C_R(2\pi)^{2-\epsilon/2}e^{i\pi \epsilon}\int_0^{\pi} dy(\sin ^2y)^{-1+
\epsilon /2} \int_y^{2\pi -y} dx [a^2\sin ^2 x - b^2 \cos ^2 x +i\eta]^
{{{\epsilon}\over 2}} $$
$$+2g^2C_R (2\pi  )^{2-\epsilon /2} e^{i\pi \epsilon} (a^2 -b^2)\int_0^{\pi}
dx [a^2 \sin ^2 x -b^2 \cos^2 x +i\eta]^{-1+\epsilon /2} $$
$$ \times \{\int_0^x dy(\sin^2 y)^{{{\epsilon}\over2}} + \int_0^{\pi-x}
dy(\sin^2y)^{{{\epsilon}\over2}}\}, \eqno(21) $$
where $u-v=2y,~u+v =2x$
The first integrand in (21) is singular at $y=0$,
where the two points $u$ and $v$ coincide. This singularity is expected
and occurs also in Euclidean space. It is
a `linear divergence'. With dimensional regularization, it
gives rise to no pole at $\epsilon =0$. The second integral in (21)
is singular at $\tan ^2 x = {b^2 \over a^2}$. This is the case
when the points $u$ and $v$ are on each others light cone, and is
the possible divergence we wish to investigate. We will see that (in
dimensional regularization at least) this singularity does not give rise to a pole at $\epsilon =0$.

Let us consider the two lines in (21), starting with the first. If we changed the order of
integration and did the $y$-integral first, we would get no pole at 
$\epsilon =0$. Therefore we may set $\epsilon =0$ in the $x$-integral.
This integral is then trivial, and, after symmetrization, we obtain
just the integral
$$\int_0^{\pi} dy (\sin ^2 y)^{-1+\epsilon/2}=
{{\Gamma({{\epsilon-1}\over2})\Gamma({1\over2})}\over{
 \Gamma ({{\epsilon}\over2})}}, \eqno(22)$$
which gives zero in the limit $\epsilon \rightarrow 0$.

Now turn to the second line in (21). Once again, we set $\epsilon =0$
in the nonsingular $y$-integral. In this case, the
$x$-integral too is well-defined for $\epsilon =0$, and is then just
$$(a^2-b^2)\int_0^{\pi} dx [a^2 \sin ^2 x -b^2 \cos ^2 x +i\eta]^{-1}
=(a^2 -b^2 )\int_{-\infty}^{\infty} 
{dz \over a^2 z^2 -b^2+i\eta}=-i\pi{(a^2 -b^2)\over ab}.\eqno(23)$$
Thus the exitence of pairs of points with null interval causes no
divergence. In fact the Minkowski space result (23) is obtainable from
the Euclidean space one simply by the continuation $b \rightarrow ib$.
It is amusing that for the case of a circular Wilson loop ($a=b$) the
Minkowski result (23) vanishes.

\beginsection Appendix A

In this Appendix we quote from [2] the result for the integral appearing
in Fig.2(b), thus confriming the validity of the approximation
made in (7) for identifying the divergence.

The relevant integral over $k$, holding $u$ fixed, is
$$\int d^2K dv [uv-K^2]^{-1}[(u-w)(v-V'.p)-(K-P)^2]^{-1}$$ $$ =i\pi^{2-\epsilon/2}
\Gamma(\epsilon /2)w^{-1}(-p^2-i\eta)^{-\epsilon/2}[a(1-a)]^{-\epsilon /2}
\theta(a) \theta (1-a), \eqno(A1)$$
where $a=u/w$. Thus there is indeed a pole, coming from the integral
with $u$ held fixed, and it is the same as would come from the approximate form in (7).

\beginsection Appendix B

In this Appendix we treat particular examples of the graphs in Fig.2(b)
and (c), for the special case where the segment $ BB' $ becomes a straight
line. For definiteness, we have completed the loop with two further
straight sides to make a parallelogram, and inserted an extra gluon $q $
(so that the trace (1) is nonzero), as in Fig.4 and Fig.5. Let the top
and bottom sides of these parallelograms be represented by the vector
$ N $. For any vector $ k $, let
$$ k=xV+yN+K_T, ~~~
 V.K_T=N.K_T=0. \eqno(B1) $$   Let us denote
$$ u=V.k, ~~~v=N.k, $$
$$ d^4k=d^2K_Tdudv\mid (V.N)\mid ^{-1}. \eqno(B2) $$  The graph in Fig.4 is
$$ W_4=-ig^4C_G Tr(t_at_d)V_{\mu}N_{\rho}\int {{d^nk}\over{(2\pi)^n}}
{{(p-2k).V}\over{k^2(p-k)^2}} $$ $$ \times {1\over{k.V}}\{
{1\over{(p-k).V}}[e^{i(p-k).V}-1]-{1\over{p.V}}[e^{ip.V}-1]\}
{1\over{q.V}}(1-e^{iq.V})e^{iq.N}  \eqno(B3) $$
Using (A1) and performing also the $ a $ integration we obtain
$$ W_4={{g^4}\over{16\pi^2}}C_G Tr(t_at_d)V_{\mu}N_{\rho}
\Gamma({{\epsilon}\over2})(-p^2-i\eta)^{-{{\epsilon}\over2}}
{1\over{q.V}}(1-e^{iq.V})e^{iq.N} $$ $$ \times {1\over{p.V}}
\{(e^{ip.V}-1)[{\rm ci} (p.V)-\ln (p.V)-C+2]-i(e^{ip.V}+1)
[{\rm si} (p.V)+{{\pi}\over2}]\}.\eqno(B4) $$
where $ C $ is the Euler's constant and $ {\rm ci(p.V) } $
and $ {\rm si(p.V)} $ are integrated cosine and sine functions.
 The graph in Fig.5 gives
$$ W_5=-{{g^4}\over{2\pi^2}}(C_R-{1\over 2}C_G)Tr(t_at_d)V_{\mu}N_{\rho}
e^{{{i\pi\epsilon}\over2}}(\mid N^2\mid )^{-{\epsilon\over2}}
(V.N)^{\epsilon}(V.p)^{-\epsilon-1}\Gamma({\epsilon\over2}) $$ $$ \times
e^{iV.p}\{{\rm ci}(V.p)+i{\rm si}(V.p)-C+{{i\pi}\over2}-\ln (V.p)\}
\times{1\over{q.V}}(1-e^{iq.V})e^{iq.N}. \eqno(B5) $$

\beginsection  References

[1] V.S. Dotsenko and S.N. Vergeles, Nucl. Phys.{\bf B169}(1980)527;

\noindent
[2] A. Andra\v si and J.C. Taylor, Nucl. Phys.{\bf B350}(1991)73;

\noindent
[3] R. Brandt, F. Neri and Masa-aki Sato, Phys. Rev.{\bf D24}(1981)879;

\noindent
[4] I.A. Korchemskaya and G.P. Korchemsky, Phys. Lett.{\bf B287}(1992)169;

\noindent
[5] A. Bassetto, I.A. Korchemskaya, G.P. Korchemsky and G. Nardelli,

\line{\hskip 0.5 cm Nucl. Phys.{\bf B408}(1993)62; \hfill}

\noindent
[6] A.M. Polyakov, Nucl. Phys.{\bf B164}(1980)171;

\noindent
[7] G.P. Korchemsky and A.V. Radyushkin, Nucl. Phys.{\bf B283}(1987)342;

\noindent
[8] A. Andra\v si and J.C. Taylor, Nucl. Phys.{\bf B323}(1989)393

\vskip 1cm

\beginsection Figure Captions

Fig.1. Graphs contributing to the charge renormalization for
general Wilson loops.

Fig.2. A portion of the Wilson loop which contains a straight lightlike
segment $ AB $ and the neighbouring portions $ A'A $ and  $ BB' $ of the
loop. Besides the graph in Fig.2(b), the graphs where a gluon is
attached to the neighbouring portion $ BB' $ contribute to the charge
renormalization. The remainder of the Wilson loop is not drawn. There
may be additional gluons attached to it.

Fig.3. The graphs to which a part of Fig.(2b) reduces after one
parameter integration.

Figs.4 and 5. Examples of complete Wilson loop with two sides along the
lightlike vector $ V $ and two sides  along vector $ N $.

\bye